Rapid ortho-to-para nuclear spin conversion of $H_2$ on a silicate dust surface

M. Tsuge[1], T. Namiyoshi[1], K. Furuya[2], T. Yamazaki[1], A. Kouchi[1], N. Watanabe[1*]

[1]Institute of Low Temperature Science, Hokkaido University, Sapporo, Hokkaido 060-0819, Japan

[2]National Astronomical Observatory of Japan, Osawa 2-21-1, Mitaka, Tokyo 181-8588, Japan

*e-mail: watanabe@lowtem.hokudai.ac.jp

Abstract

The $H_2$ molecule has two nuclear spin isomers, the so-called ortho and para isomers. Nuclear spin conversion (NSC) between these states is forbidden in the gas phase. The energy difference between the lowest ortho and para states is as large as 14.7 meV, corresponding to ~170 K. Therefore, each state of $H_2$ differently affects not only the chemistry but also the macroscopic gas dynamics in space, and thus, the ortho-to-para abundance ratio (OPR) of $H_2$ has significant impacts on various astronomical phenomena. For a long time, the OPR of nascent $H_2$ upon formation on dust grains has been assumed to have a statistical value of three and to gradually equilibrate in the gas phase at the temperature of the circumstances. Recently, NSC of $H_2$ was experimentally revealed to occur on water ice at very low temperatures and thus incorporated into gas-dust chemical models. However, $H_2$ molecules should form well before dust grains are coated by water ice. Information about how the OPR of $H_2$ behaves on bare silicate dust before ice-mantle formation is lacking. Knowing the influence of the OPR of $H_2$ if the OPR changes even on a bare silicate surface within an astronomically meaningful time scale is desirable. We report the first laboratory measurements of NSC of $H_2$ physisorbed on amorphous silicate ($Mg_2SiO_4$) at temperatures up to 18 K. The conversion was found to occur very rapidly.

Unified Astronomy Thesaurus concepts: Astrochemistry (75); Molecular clouds (1072); Dense interstellar clouds (371); Interstellar molecules (849); Interstellar dust (836); Laboratory astrophysics (2004);

# 1. Introduction

The $H_2$ molecule consists of two protons and thus has a total nuclear spin of 0 or 1 with multiplicities of singlet and triplet, respectively. According to the Pauli exclusion principle, the nuclear spin states of 1 (ortho) and 0 (para) are only coupled with the rotational states of odd and even quantum numbers, respectively. In thermal equilibrium, the ortho-to-para abundance ratio (OPR) is expressed by the ratio of rotational Boltzmann distributions for odd and even rotational numbers with spin degeneracies. At temperatures above approximately 200 K, the OPR almost reaches a statistical value of three, while at the low-temperature limit, it becomes zero. The energetically lowest ortho state with rotational number $J = 1$ lies approximately 14.7 meV above the para state with $J = 0$. In the gas phase, radiative transition between these two states is forbidden and, thus, nuclear spin conversion (NSC) only occurs via spin exchange reactions with protons or hydrogen atoms. The conversion timescale due to these processes in molecular clouds (MCs) is as slow as approximately $10^5$–$10^7$ yr (Wilgenbus et al. 2000; Flower, Pineau Des Forêts, & Walmsley 2006). Therefore, $H_2$ molecules in each state can be considered energetically different species. Because of the large energy gap between the ortho and para states, the OPR of $H_2$ has important astronomical meaning. For example, the OPR affects the gas dynamics of core formation in star-forming regions because the heat capacity of $H_2$ gas is different between the ortho and para states (Vaytet, Tomida, & Chabrier 2014). Because of the slow conversion between two spin isomers, the OPR of $H_2$ has been used as a tracer of the age of clouds (Pagani et al. 2013). In cold (~10 K) regions such as an MC, the OPR can control the chemical evolution. One of the most discussed issues is the effect of the OPR on the deuterium fractionation of molecules in the gas phase (Bovino et al. 2017). The ionic species $H_3^+$ is involved in the formation of various interstellar molecules. Once $H_3^+$ is deuterated by an ion-molecule reaction, $H_3^+ + HD \rightarrow H_2D^+ + H_2$, it works as a dominant deuterium fractionation pathway in MCs. When reactant $H_3^+$ and products $H_2D^+$ and $H_2$ are all in the para state, this reaction is exothermic by 230 K. However, if $H_2$ is in the ortho state, then the exothermicity is significantly reduced, resulting in the occurrence of a reverse process that destroys $H_2D^+$ even in MCs.

In contrast to processes in the gas phase, little is known about how the OPR of $H_2$ behaves on dust grains. $H_2$ has been widely accepted to predominantly form via H-H recombination on the surface of dust grains (Wakelam et al. 2017). Furthermore, $H_2$ molecules in the gas phase inevitably collide and interact with dust grains. Nevertheless, processes on dust surfaces concerning the OPR of $H_2$ remained unclear for a long time because of the difficulties in both experimental and theoretical approaches. Recently, several experiments finally shed light on the behavior of the $H_2$ OPR on ice surfaces (Sugimoto & Fukutani 2011; Ueta et al. 2016). The OPR of nascent $H_2$ formed by H-H recombination was revealed to be almost 3, and NSC can occur on water ice within approximately $10^3$ s depending on the temperature (Ueta et al. 2016; Watanabe et al. 2010). These experimental findings have now been incorporated into gas-dust chemical models (Bovino et al. 2017; Furuya et al. 2019). However, $H_2$ formation should be activated and equilibrate on base silicate or carbonaceous dust well before ice-mantle formation. Therefore, knowing whether NSC occurs to change the OPR of $H_2$ on these surfaces is highly desirable.

## 2. Experiment

The experiments were performed in an ultrahigh vacuum chamber (~$10^{-8}$ Pa) with a doubly differentially pumped molecular beam source. An amorphous silicate film was deposited on an aluminum substrate by pulsed laser ablation of an $Mg_2SiO_4$ target in situ; the preparation and characterization of the film are described in Appendix (section A.1). The substrate was located at the center of the main chamber and could be cooled by a closed-cycle helium refrigerator.

The NSC of $H_2$ on the $Mg_2SiO_4$ surface was investigated with a combination of temperature programmed desorption (TPD) and resonance-enhanced multiphoton ionization (REMPI) techniques: the TPD-REMPI method. The timing chart for TPD-REMPI experiments is shown in Figure 1. The temperature of the substrate was measured with a silicon diode and controlled with a fluctuation of less than 0.5 K. Prior to each measurement, the substrate was warmed to 55 K to remove $O_2$ molecules from the surface since $O_2$ molecules are known to enhance the ortho-to-para conversion of $H_2$. At temperatures of 10–18 K, $H_2$ molecules with an OPR of 3 were deposited on the substrate as a pulsed molecular beam (~300 μs pulse duration, 200 Hz, 3000 shots) with an incident angle of 45° with respect to the surface normal. The pulsed $H_2$ beam was produced by expansion of normal $H_2$ gas (300 kPa) through a pulsed valve (100 μm-diameter orifice), skimmed by a skimmer, and introduced into the main chamber through an orifice of 1−2 mm. By using an $H_2$ molecular beam instead of a continuous beam, the deposition duration is significantly reduced so that the ortho-to-para conversion during deposition is suppressed. An increase in undesired background $H_2$ in the chamber is also avoided. Therefore, the use of pulsed molecular beam suppresses NSC due to gas-phase secondary processes and NSC on the coldhead (i.e., metal surfaces). At 10 K, the $H_2$ dose was approximately $3 \times 10^{14}$ molecules cm$^{-2}$, corresponding to an H$_2$ coverage of approximately 0.03, at which the interaction between H$_2$ molecules would be negligible. After a certain waiting time, $t_{w.t.}$, of 10–810 s, the sample was warmed to 55 K at a ramp rate of 20 K min$^{-1}$: the TPD method. The desorbing H$_2$ molecules were ionized by the REMPI method and detected with a time-of-flight spectrometer. Laser radiation in a wavelength range of 201–203 nm was provided by an Nd$^{3+}$:YAG laser-pumped dye laser with subsequent frequency doubling and mixing in potassium dihydrogen phosphate and beta barium borate crystals, respectively. In this wavelength range, H$_2$ molecules can be rotational-state-selectively ionized by (2 + 1) REMPI via the $E, F^1(v' = 0, J' = J'') \leftarrow X^1$ ($v'' = 0, J'' = 0$ or 1) transition. The REMPI laser was focused by a plano-convex lens with $f$ = 300 mm approximately 1.0 mm above the sample surface. In this paper, we call the H$_2$ signal detected during a TPD run a TPD-REMPI signal. At low temperatures, *ortho*-H$_2$ (*o*-H$_2$) and *para*-H$_2$ (*p*-H$_2$) are in the $J = 1$ and $J = 0$ states, respectively, and thus, we measured TPD-REMPI signals for these states.

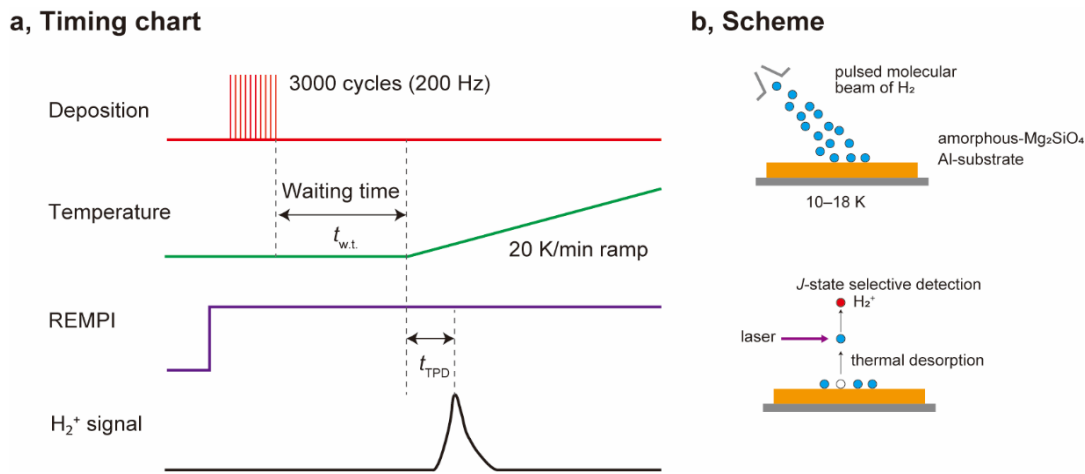

Figure 1. Schematic of the TPD-REMPI experiment. a, Timing chart of the TPD-REMPI experiment. A series of molecular beam pulses of normal $H_2$ are first deposited on an $Mg_2SiO_4$ sample maintained at 10–18 K. After a certain waiting time ($t_{w.t.}$, 10–810 s), the temperature of the substrate is raised at a 20 K min$^{-1}$ ramping rate; this process is the so-called temperature-programmed desorption (TPD) process. The laser for REMPI is always operating during each run with a 10 Hz repetition rate and a pulse energy of 200 μJ pulse$^{-1}$. The $H_2$ molecules desorbed during a TPD run, i.e., at $t = t_{w.t.} + t_{TPD}$, are rotational-state-selectively ionized to $H_2^+$ by the REMPI laser and detected with a linear time-of-flight spectrometer. b, Schematic for the $H_2$ deposition (top) and TPD-REMPI (bottom) processes.

## 3. Results and Discussion

### *3.1 Determination of NSC Time Constants*

The experimental results obtained at 10 K are presented in Figure 2. Figure 2a shows a series of TPD-REMPI data measured for $t_{w.t.}$ = 10, 410, and 810 s. The growth of the $J = 0$ signal and decay of the $J = 1$ signal with increasing $t_{w.t.}$ are readily recognized. The time variation of integrated TPD-REMPI signals is plotted in Figure 2b. The sum of the $J = 0$ and $J = 1$ signals is constant within the experimental error, ensuring that accumulation of background $H_2$ during the waiting time and desorption upon NSC are negligible. Consequently, the decay of the $J = 1$ signal is attributed to the ortho-to-para conversion.

When the temperature of the substrate is sufficiently low such that thermal desorption can be ignored, the number density of $o$-$H_2$ ($J = 1$) decreases with time due to ortho-to-para conversion, while that of $p$-$H_2$ ($J = 0$) increases. This situation is applicable to temperatures of 10, 12, and 14 K in the present experiments. Thus, if one monitors the number density, $[o\text{-}H_2]_t$, it shows a single exponential decay

$$[o\text{-}H_2]_t = [o\text{-}H_2]_0 \exp(-k_{OP}t), \tag{1}$$

where $k_{OP}$ is the ortho-to-para conversion rate constant and $[o\text{-}H_2]_t + [p\text{-}H_2]_t = [o\text{-}H_2]_0 + [p\text{-}H_2]_0$ = const. Fitting of the decay of the $J = 1$ TPD-REMPI signal by a single exponential function gives a decay time constant ($1/k_{OP}$) of 980 ± 90 s at 10 K. The OPR, ~1.8, for $t_{w.t.}$ = 10 s is already smaller than the statistical value of 3 because NSC proceeds even during the TPD run. As detailed in section A.2, this enhancement has little effect on the obtained time constants.

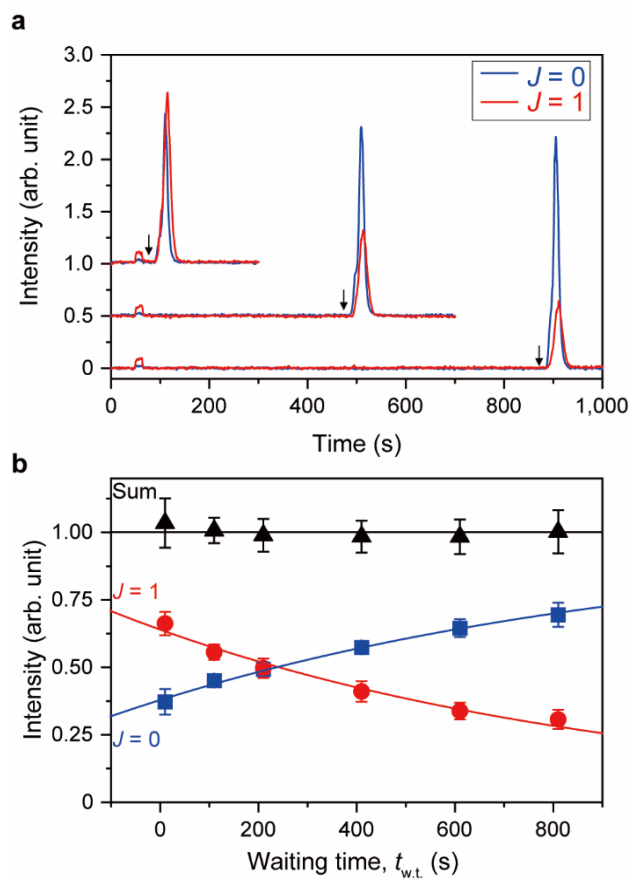

Figure 2. Time evolution of $o$- and $p$-$H_2$ populations at 10 K. a, $H_2$ intensities for $J = 0$ (blue) and $J = 1$ (red) detected by the REMPI method at different waiting times (from top to bottom, 10, 410, and 810 s). The weak signals at 50–65 s are due to gaseous $H_2$ supplied from the molecular beam source. The starting time of each TPD run is indicated by vertical arrows. b, Time evolution of TPD-REMPI signals for $J = 0$ (blue square), $J = 1$ (red circle), and their sum (black triangle). The intensities were obtained by integrating the TPD-REMPI signal. The error bars represent the statistical errors among 5 respective measurements. Solid curves are the results of single exponential fitting.

The time variation of TPD-REMPI signals for temperatures of 12, 14, 16, and 18 K are presented in Figure 3. As shown in Figures 3a and 3b, the sum of the $J = 0$ and $J = 1$ TPD-REMPI signals is constant at 12 and 14 K, similar to the case at 10 K. Thus, the decay of $J = 1$ TPD-REMPI signal can be fitted by Equation (1) to obtain decay time constants of $560 \pm 60$ and $360 \pm 20$ s for 12 and 14 K, respectively.

At 16 and 18 K (Figures 3c and 3d), decay of the sum as a function of $t_{w.t.}$ is observed. This decay is presumably due to thermal desorption during the waiting time. In this case, differential rate equations for the surface number densities of $o$-$H_2$ and $p$-$H_2$ are written as follows:

$$\frac{d[o\text{-}H_2]}{dt} = -k_{sub}[o\text{-}H_2] - k_{OP}[o\text{-}H_2], \tag{2}$$

$$\frac{d[p\text{-}H_2]}{dt} = -k_{sub}[p\text{-}H_2] + k_{OP}[o\text{-}H_2], \tag{3}$$

where $k_{sub}$ is the rate constant for unimolecular sublimation, which is assumed to be the same for $o$-$H_2$ and $p$-$H_2$; in reality, $k_{sub}$ for $o$-$H_2$ might be smaller because of a slightly large binding energy, as deduced from the TPD spectra. Equation (2) is readily integrated to become

$$[o\text{-}H_2] = [o\text{-}H_2]_0 e^{-(k_{sub}+k_{OP})t}. \tag{4}$$

An integration of the sum of Equations (2) and (3) gives the following equation for the total $H_2$ on the surface:

$$[H_2]_t = [o\text{-}H_2]_t + [p\text{-}H_2]_t = [H_2]_0 e^{-k_{sub}t}. \tag{5}$$

Therefore, the sublimation rate constant ($k_{sub}$) is obtained from single exponential fitting of the decay of $[H_2]_t$ (i.e., the sum of $[o\text{-}H_2]_t$ and $[p\text{-}H_2]_t$), and $k_{sub} + k_{OP}$ is obtained from fitting of the decay of $[o\text{-}H_2]_t$. Consequently, the ortho-to-para conversion time constant, $1/k_{OP}$, is determined. Accordingly, the time constant for NSC was extracted from the fitting of $J = 1$ and the sum to be $290 \pm 20$ and $260 \pm 70$ s for 16 and 18 K, respectively, whereas the time constants for sublimation ($1/k_{sub}$) were $3300 \pm 200$ and $400 \pm 80$ s.

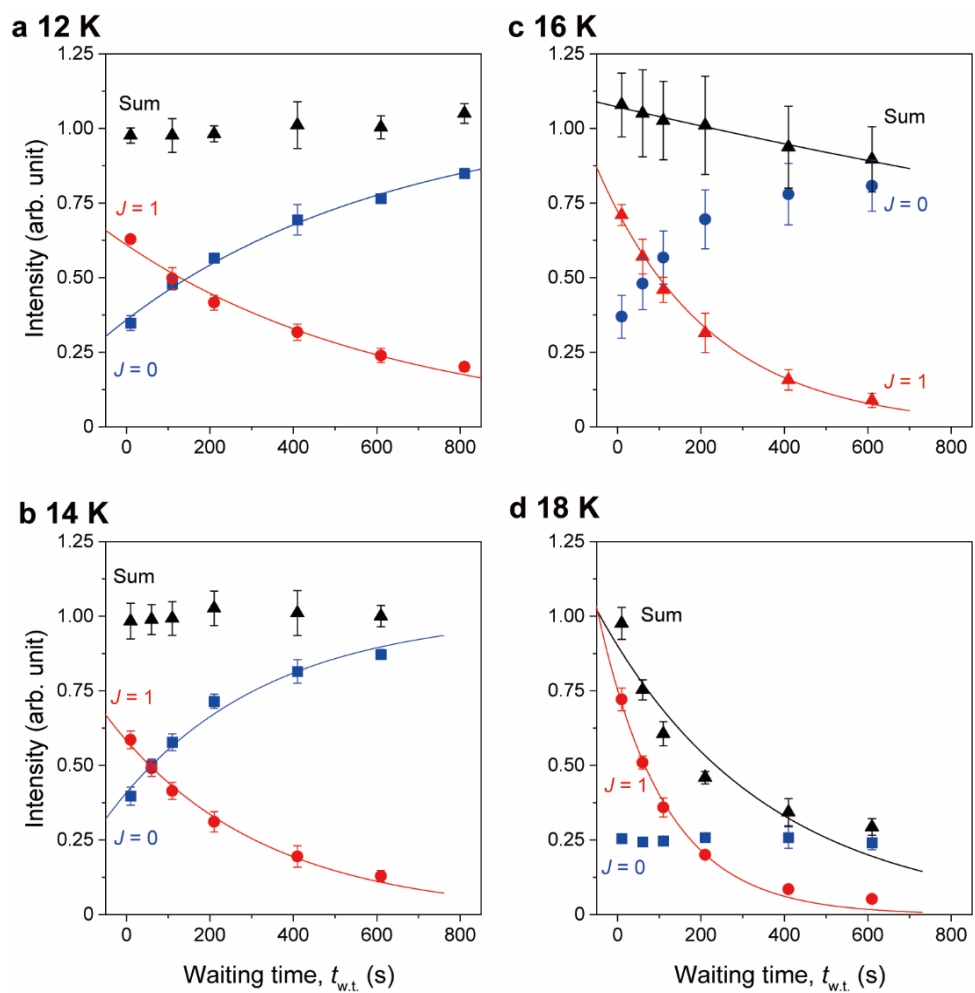

Figure 3. Time evolution of TPD-REMPI intensities for $J = 0$, $J = 1$, and their sum. a, b, c, and d, Time evolution of TPD-REMPI signals for $J = 0$ (blue square), $J = 1$ (red circle), and their sum (black triangle) at temperatures of 12, 14, 16, and 18 K, respectively. The intensities were obtained by integrating the TPD-REMPI signal. The error bars represent the statistical errors among 5 respective measurements. Solid curves are the results of single exponential fitting.

*3.2 Comparison of NSC Time Constants*

The NSC time constants determined in this work for an amorphous $Mg_2SiO_4$ surface are plotted as a function of substrate temperature in Figure 4, together with those reported for an amorphous solid water (ASW) surface (Ueta et al. 2016). These time constants are listed in Table 1. The time constants for the $Mg_2SiO_4$ surface are smaller than those for the ASW surface by a factor of 2–3. In the case of ASW, the time constant significantly decreases in the 9.2 to 12 K region and remains constant above 12 K. In the case of $Mg_2SiO_4$, we observe a gradual decrease in time constant as a function of temperature in the investigated temperature range (10–18 K). Because of fast unimolecular sublimation, we were unable to determine the time constant above 18 K. However, an analysis described in section A.2 indicates that the time constant would not be smaller than 200 s even at the sublimation temperature (20–26 K) in the TPD runs. For comparison, we measured the NSC of $D_2$ at 18 K (section A.3). The time constant is much larger than that of $H_2$ at 18 K, consistent with previous experiments on ASW (Sugimoto & Fukutani 2011; Ueta et al. 2016).

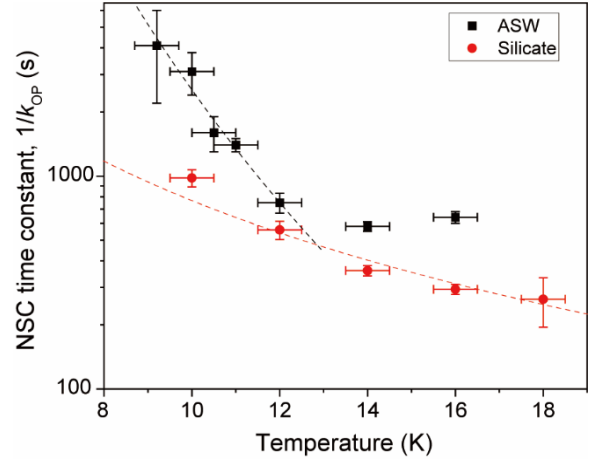

Figure 4. Ortho-to-para conversion time constants on silicate (amorphous $Mg_2SiO_4$) and ASW. The temperature dependences of the ortho-to-para conversion time constants ($1/k_{OP}$) for silicate (this work) and ASW (Ueta et al. 2016) are plotted. Vertical error bars represent the errors originating from the decay curve fitting. Horizontal error bars represent the temperature fluctuation in the experiments. Dashed lines represent the results of fitting assuming the $k_{OP} \propto T^n$ relation (Scott & Jeffries 1962). For amorphous silicate, $n = 1.9 \pm 0.3$ was obtained for the temperature range 10–18 K, while, for ASW, $n = 7.1 \pm 0.6$ for the range 9.2–12 K. Time constants are summarized in Table 1.

Table 1. NSC time constants determined for $H_2$ on amorphous $Mg_2SiO_4$ and ASW surfaces at various temperatures.

| Temperature (K) | NSC time constant (s) | |
| --- | --- | --- |
| | $Mg_2SiO_4$ | ASW[a] |
| 9.2 | | 4,100 ± 1,900 |
| 10 | 980 ± 90 | 3,100 ± 700 |
| 10.5 | | 1,600 ± 300 |
| 11 | | 1,400 ± 100 |
| 12 | 560 ± 60 | 750 ± 80 |
| 14 | 360 ± 20 | 580 ± 30 |
| 16 | 290 ± 20 | 640 ± 40 |
| 18 | 260 ± 70 | |

[a]Ueta et al. 2016.

*3.3 Temperature-dependence of NSC Rate*

The ortho-to-para conversion of $H_2$ on surface involves the NSC and subsequent energy dissipation processes. The presence of magnetic field is known to enhance the NSC process (Fukutani & Sugimoto 2011). Because an $Mg_2SiO_4$ film does not have magnetic moment and an adsorption of paramagnetic $O_2$ molecules is avoided, such an enhancement would be negligible in the present experiment. However, we cannot exclude magnetic moment induced by defects near the sample surface. To discuss the effect of such magnetic moments, an extensive surface characterization is required. The mechanisms of NSC on nonmagnetic surface have been theoretically discussed in recent years but those are not yet conclusive (Fukutai & Sugimoto 2011; Ilisca 2018; Ilisca & Ghiglieno 2014). Clarifying it is out of the scope of this paper. Here, we focus our discussion on the mechanism of energy dissipation process required for the NSC. The energy gap between the $J = 1$ and $J = 0$ states is 14.7 meV ($\approx$170 K) for the isolated $H_2$ but may become smaller due to the suppression of rotational motion for $H_2$ adsorbates. Therefore, the corresponding energy should be released into the solid. The presence of temperature dependence (Figure 4) indicates that phonons play a role in the energy dissipation process.

Spin-lattice relaxation will occur through one-phonon or two-phonon processes. In the one-phonon process, the excess energy associated with the NSC process is transferred to the surroundings (i.e., phonon excitation in the thermal bath). Depending on the energy matching condition between the excess energy and phonon modes of the surroundings, the one-phonon process shows two distinct temperature dependences (Scott & Jeffries 1962). When there is a satisfactory match between them, the one-phonon process is known to show a temperature dependence of NSC rate as $k_{\text{OP}}(T) =$ $A \times \coth(\frac{\delta}{2T})$, where $T$ is the solid temperature and $\delta$ is the energy gap between these two spin states. On the other hand, when there is a mismatch between them, the process becomes a slower phonon-limited "bottle-neck" process. This process shows a distinct temperature dependence, $k_{\mathrm{OP}}(T) = A \times T^n$ ($n \sim 2$).

The two-phonon process proceeds via the simultaneous absorption (from the initial state ($J = 1$) to the intermediate state) and emission (from the intermediate state to the final state ($J = 0$)) of phonons. Based on whether the intermediate state is real (i.e., $J = 2$ state) or virtual, the process is classified into Orbach process or Raman process, respectively. The NSC rate of Orbach process follows $k_{\mathrm{OP}}(T) = A \times \exp(-\frac{\Delta}{T})$, where $\Delta$ corresponds to the energy gap between the initial and intermediate states (Scott & Jeffries 1962). On the other hand, the Raman process shows a power law dependence, $k_{\mathrm{OP}}(T) = A \times T^n$ ($n = 7$ or $9$). For example, Ueta et al. (2016) attributed the temperature dependence observed for $H_2$ NSC on ASW to this process (Figure 4).

Among the phonon processes described above, the one phonon process expressed by $k_{\mathrm{OP}}(T) = A \times \coth(\frac{\delta}{2T})$ can be excluded because it becomes almost temperature independent in the present experimental condition, where $\delta = 170$ K and $10$ K $\leq T \leq 18$ K. The Orbach process ($k_{\mathrm{OP}}(T) = A \times \exp(-\frac{\Delta}{T})$) with $\Delta = 399$ K that is the energy gap between $J = 1$ and $J = 2$ (Silvera 1980) was found not to reproduce the observed temperature dependence. Then, we fitted the experimental data by the expression $k_{\mathrm{OP}}(T) = A \times T^n$, which gave $n = 1.9 \pm 0.3$ (see Figure 4). As a result, we attribute the observed temperature dependence to a slower phonon-limited "bottle-neck" process, where one-phonon process dominates over two-phonon Orbach and Raman processes and there is a mismatch between the excess energy associated with the NSC process and phonon modes of the thermal bath. It is not easy to further discuss the energy matching condition because the exact energy gap for $H_2$ on $Mg_2SiO_4$ surface and the surface-phonon frequencies of amorphous $Mg_2SiO_4$ are not known. Nonetheless, an inelastic neutron scattering study of crystalline $Mg_2SiO_4$ (Ghose et al. 1987) showed that an $Mg_2SiO_4$ solid possesses phonon modes in the range 10−40 meV, indicating that the excess energy associated with the direct NSC process (≤14.7 meV) can be eventually dissipated into the $Mg_2SiO_4$ bath. We note that this will not be the sole mechanism that explain a rapid NSC of $H_2$ on $Mg_2SiO_4$ surface, since the NSC of molecules on surface is complicated; e.g., several mechanisms have been suggested to explain the observed $H_2$ NSC rates on ASW (Fukutani & Sugimoto 2011, Ueta et al. 2016, Ilisca 2018).

*3.4 Efficiency of ortho-to-para Conversion under Astrophysical Environments*

In star-forming regions, almost all hydrogen is present as $H_2$ in the gas phase rather than on grain surfaces because of the very low $H_2$-$H_2$ surface binding energy (~100 K (Lee, Gowland, & Reddish 1971)), and because of the limited number of binding sites per $H_2$ molecule (~$10^{-4}$ site per $H_2$). Here, we derive the rate (cm$^{-3}$ s$^{-1}$) at which NSC on grain surfaces affects the gas-phase $H_2$ OPR under astrophysical environments using the experimentally derived $k_{OP}$. We denote the rate as $R^{surf}_{\alpha\to\beta}$, where α and β indicate either ortho or para. $R^{surf}_{\alpha\to\beta}$ depends on the (i) collision rate of gaseous $H_2$ with grain surfaces ($R_{col}$), (ii) sticking probability ($S$), (iii) ortho-to-para conversion time constant, and (iv) sublimation timescale of adsorbed $H_2$ and can be described as (Bovino et al. 2017; Fukutani & Sugimoto 2013; Furuya et al. 2019)

$$R^{surf}_{\alpha\to\beta} = \eta_{\alpha\to\beta}(1-\Theta)SR_{col}(\alpha\text{-}H_2), \qquad (6)$$

where Θ is the fraction of adsorption sites occupied by $H_2$. We consider the factor $(1 − Θ)$ in Equation (6), assuming that only one $H_2$ molecule is allowed to be adsorbed per adsorption site. Then, the product $(1 − Θ)SR_{col}(α\text{-}H_2)$ expresses the adsorption rate of $H_2$. The term $\eta_{o \to p}$ ($\eta_{p \to o}$) represents the yield of gaseous $o$-$H_2$ ($p$-$H_2$) per $p$-$H_2$ ($o$-$H_2$) adsorption, which depends on the competition between NSC and sublimation of adsorbed $H_2$. To determine $\eta_{o \to p}$ under astrophysical environments, we basically follow the procedure developed in our previous work, in which NSC of $H_2$ on an ASW surface was studied (Furuya et al. 2019).

The surface coverage of $H_2$ is expected to be in adsorption-desorption equilibrium since the surface adsorption site density is $\sim 10^{16}$ site cm$^{-2}$ and the $H_2$ fluence (the time integral of the $H_2$ flux) on grain surfaces reaches this value in a very short timescale ($\sim 10$ ($10^3$ cm$^{-3}$/$n(H_2)$) yr) compared to the dynamical timescale (i.e., lifetime of MCs). Under adsorption-desorption equilibrium, the $H_2$ OPR of the gas desorbing from the surface can be expressed as

$$R_{sub}(o\text{-}H_2)/R_{sub}(p\text{-}H_2) = [(1 − \eta_{o \to p})f_o + \eta_{p \to o}f_p]/[\eta_{o \to p}f_o + (1 − \eta_{p \to o})f_p], \quad (7)$$

where $R_{sub}$ is the sublimation rate of $H_2$. $f_o$ and $f_p$ are the fractions of $o$-$H_2$ and $p$-$H_2$, respectively, in adsorbing $H_2$ (or equivalently $H_2$ in the gas phase). With the relation $\eta_{p \to o} = \eta_{o \to p}\gamma$, where $\gamma$ is the thermalized value of the $H_2$ OPR, one can obtain

$$\eta_{o \to p} = [f_o − f_p R_{sub}(o\text{-}H_2)/R_{sub}(p\text{-}H_2)]/[(f_o − \gamma f_p)(1 + R_{sub}(o\text{-}H_2)/R_{sub}(p\text{-}H_2))]. \quad (8)$$

We quantify $R_{sub}(o\text{-}H_2)$ and $R_{sub}(p\text{-}H_2)$ using the numerical simulations described below, followed by evaluation of $\eta_{o \to p}$ and $\eta_{p \to o}$ from Equation (8).

We numerically solve a set of rate equations for $o$-$H_2$ and $p$-$H_2$ that describe the adsorption of gaseous $H_2$, thermal desorption, thermal hopping, and NSC of adsorbed $H_2$, considering various adsorption sites with different potential energy depths on surfaces (see Equations 6 and 7 in Furuya et al. 2019). Regarding chemical species, only $o$-$H_2$ and $p$-$H_2$ in the gas phase and on grain surfaces are considered in this model. Initially, all $H_2$ is assumed to be in the gas phase with an OPR of 3, i.e., $f_o$ and $f_p$ are 0.75 and 0.25, respectively. For the binding energy distribution of $H_2$, we use the experimentally determined binding energy distribution of HD on amorphous $MgFeSiO_4$ in the literature (Figure 5) (He & Vidali 2014). The hopping-to-binding energy ratio ($\chi$) is poorly constrained, and here, we assume a conservative value of 0.8. The choice of the exact value of $\chi$ does not significantly affect our simulation results because $\eta_{o \to p}$ is close to unity even for $\chi = 0.8$ (see below) and because a lower $\chi$ tends to lead to a higher $\eta_{o \to p}$ (Furuya et al. 2019). The sticking probabilities of $H_2$ to silicate and to ASW are taken from Chaabouni et al. 2012 and He, Acharyya, & Vidali 2016, respectively.

We run a small grid of pseudo-time-dependent simulations (i.e., the gas density and temperature are fixed with time in each simulation), varying the temperature from 10 to 20 K, and evaluate $\eta_{o \to p}$ using Equation (8). For temperatures from 10 to 18 K, we adopt the measured NSC time constant or the linearly interpolated value. Above 18 K, we assume that the time constant is the same as that at 18 K. As shown in Figure 6a, $\eta_{o \to p}$ is close to unity, i.e., every $o$-$H_2$ adsorption leads to NSC, up to 18 K. $\eta_{o \to p}$ decreases at higher temperatures and is ~0.25 at 20 K. For the evaluation of $\eta_{o \to p}$, we chose the time at which the total $H_2$ fluence reaches $10^{17}$ cm$^{-2}$, corresponding to the duration time of ~100 ($10^3$ cm$^{-3}$/$n(H_2)$) yr. By this time, the $H_2$ coverage on the surface reaches adsorption-desorption equilibrium under all physical conditions explored here. We confirmed that $\eta_{o \to p}$ does not change with time once the system reaches adsorption-desorption equilibrium. For an ASW surface, $\eta_{o \to p}$ is close to unity only at surface temperatures lower than 13 K. The higher $\eta_{o \to p}$ for silicate than for ASW at a given temperature is due to (i) the smaller ortho-to-para conversion time constant, as found in this work (Figure 4), and (ii) the higher binding energy (Figure 5).

The yield of $p$-$H_2$ per $o$-$H_2$ collision with silicate (i.e., $\eta_{o \to p}(1 - \Theta)S$) is shown in Figure 6b. The yield for silicate takes the maximum value of ~0.8 at ~14–18 K. In other words, every $o$-$H_2$ collision with silicate surfaces produces approximately one $p$-$H_2$ in the temperature range between 14 K and 18 K; i.e., the timescale in which NSC on grain surfaces affects the gas-phase $H_2$ OPR ($\tau_{OP} = n(o\text{-}H_2)/R_{o \to p}^{surf}$) is comparable to the collisional timescale of $o$-$H_2$ with grain surfaces (~3 ($10^3$ cm$^{-3}$/$n_H$) Myr, where $n_H$ is the number density of hydrogen nuclei). At warmer temperatures, the residence time of $H_2$ on surfaces is too short for NSC. At lower temperatures, the exchange of gaseous and icy $H_2$ is inefficient (i.e., $\Theta$ is high; adsorbed $H_2$ does not desorb and hinders other gaseous $H_2$ from being adsorbed). At a given temperature, $\Theta$ increases with increasing $H_2$ gas density. This is the reason why the yield is slightly lower at the $H_2$ gas density of $10^4$ cm$^{-3}$ compared to that at $10^3$ cm$^{-3}$.

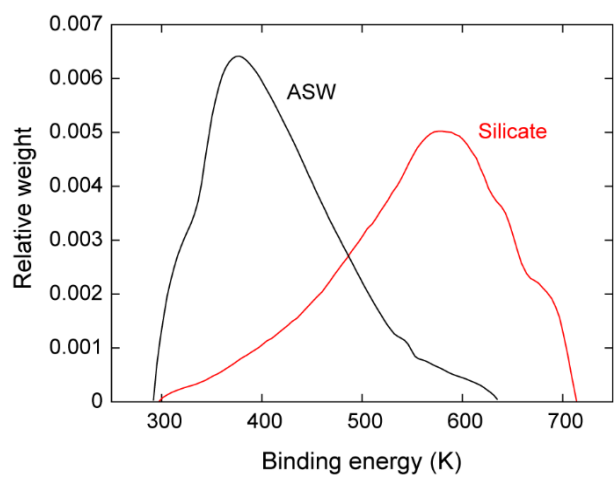

Figure 5. Binding energy distribution of $H_2$ on surfaces. The binding energy distribution of $H_2$ on silicate (red line) adopted in our numerical simulations is shown along with that on ASW (black line) for comparison. The data were taken from He & Vidali 2014.

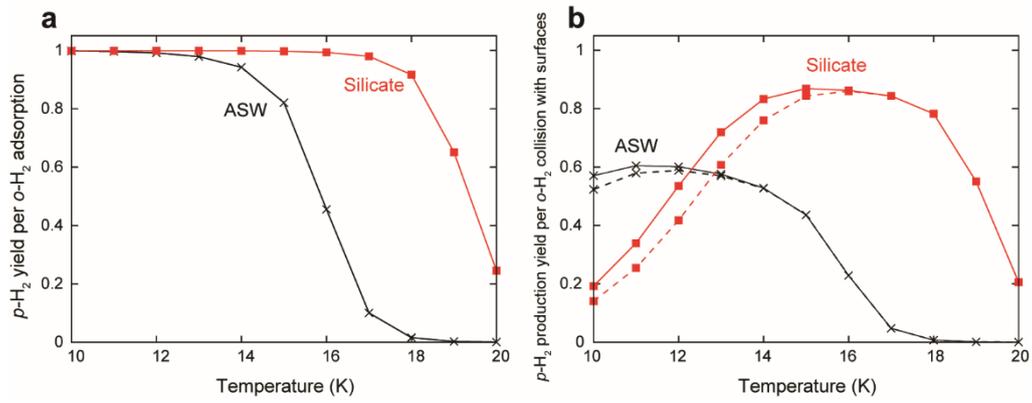

Figure 6. a, Yields of $p$-$H_2$ per $o$-$H_2$ adsorption ($\eta_{o \to p}$) as functions of temperature. The red line shows the yield for silicate (this work), while the black line shows that for ASW (Furuya et al. 2019) for comparison. The $H_2$ gas density is assumed to be $10^3$ cm$^{-3}$, but the density dependence is very weak; the yields at $10^4$ cm$^{-3}$ are almost identical to those at $10^3$ cm$^{-3}$. b, Production yields of $p$-$H_2$ per $o$-$H_2$ collision as functions of temperature. The solid lines represent the values at the $H_2$ gas density of $10^3$ cm$^{-3}$, while the dashed lines represent the values at $10^4$ cm$^{-3}$.

*3.5 Astrophysical Implication*

In the ISM, H$_2$ predominantly forms via recombination of two H atoms on grain surfaces and desorbs into the gas phase. The conversion of H atoms into H$_2$ molecules finishes before dust grains are coated by ice mantles (e.g., Hocuk & Cazaux 2015; Furuya et al. 2015). The OPR of H$_2$ upon formation on the surfaces is three (Watanabe et al. 2010). The dust temperature observed in the diffuse interstellar medium (ISM) is ~20 K, and it is lower in MCs (15–17 K) (Planck Collaboration et al. 2014). We expect that NSC of H$_2$ on bare grain surfaces affects the H$_2$ OPR from the formation stage of H$_2$ (i.e., in MCs); the NSC rate ($\tau_{OP}^{-1} n(o\text{-H}_2)$) should be comparable or greater than the H$_2$ formation rate on grains ($R_{\text{form}}(\text{H}_2)$) when $n(o\text{-H}_2) \gtrsim n(\text{atomic H})$ because the H$_2$ formation rate is given by half the adsorption rate of atomic H on grain surfaces, while the NSC rate is very close to the collisional rate of $o$-H$_2$ to the grain surfaces.

To confirm this, we consider a simple model of the evolution of $o$-H$_2$ and $p$-H$_2$ in the gas phase, where only the H$_2$ formation on grain surfaces and NSC of H$_2$ on grain surfaces are taken into account:

$$\frac{dn(o\text{-H}_2)}{dt} = b_o R_{\text{form}}(\text{H}_2) - R_{o \to p}^{surf} + R_{p \to o}^{surf}, \qquad (9)$$

$$\frac{dn(p\text{-H}_2)}{dt} = b_p R_{\text{form}}(\text{H}_2) + R_{o \to p}^{surf} - R_{p \to o}^{surf}, \qquad (10)$$

$$\frac{dn(\text{atomic H})}{dt} = -2 R_{\text{form}}(\text{H}_2), \qquad (11)$$

where $b_o$ and $b_p$ are the branching ratios to form $o$-H$_2$ and $p$-H$_2$, respectively, for H$_2$ formation on grain surfaces. We solve the above ordinary differential equations under the physical conditions appropriate for MCs, $n_H = 10^3$ cm$^{-3}$ and the gas and dust temperatures of 15 K. It should be noted that our model does not consider the formation of ice layers and we assume that dust grains are always bare. In reality, dust grains are covered by ice at some point, and after that the NSC on bare grains is no more relevant. In the ISM, two types of dust grains exist: silicate and carbonaceous materials. We treat all bare grains as silicate in this model, and the total cross section per H nuclei is set to $1 \times 10^{-21}$ cm$^{-2}$. The NSC on amorphous diamond-like carbon surfaces was observed in our previous experiments (Tsuge et al. 2019), but the time constant was not derived. The time evolution of the atomic H and H$_2$ abundances and the H$_2$ OPR are shown in Figure 7. As expected, when $n(H_2) \sim n$(atomic H), the H$_2$ OPR is already lower than three. Since the density dependence of $R_{o \to p}^{surf}$ and $R_{form}(H_2)$ is the same, this result does not depend on the chosen density. The results indicate that NSC of H$_2$ on silicate surfaces is rapid enough to reduce the H$_2$ OPR to be lower than three when the conversion of H atoms into H$_2$ occurs in the ISM.

NSC of H$_2$ also occurs in the gas phase via proton exchange reactions with H$^+$ and H$_3^+$ (Honvault et al. 2011; Hugo, Asvany, & Schlemmer 2009). After dust grains are covered by ice, NSC of H$_2$ occurs at least on ASW (Watanabe et al. 2010). In order to reveal the contribution of each process to the evolution of H$_2$ OPR, astrochemical simulations of the MC formation with the NSC processes of H$_2$ both on grain surfaces and in the gas phase are demanded. Such studies are postponed for future work.

This work was partly supported by JSPS Grant-in-Aid for Specially promoted Research (JP17H06087), Grant-in-Aid for Scientific Research (C) (JP18K03717), and Grant-in-Aid for Early-Career Scientists (JP17K14245)

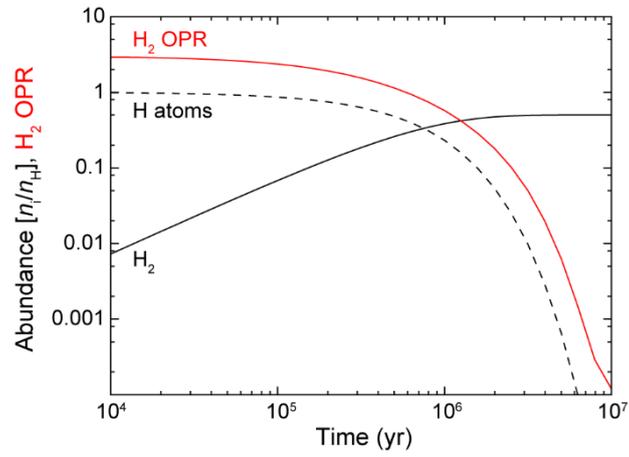

Figure 7. The time evolution of the atomic H (dashed black line) and $H_2$ (solid black line) abundances and the $H_2$ OPR (red line) predicted by our simple model (see the main text). The density of hydrogen nuclei ($n_H$) is set to $10^3$ cm$^{-3}$ and the gas and dust temperatures are set to 15 K.

# Appendix

*A.1. Preparation and Characterization of the Amorphous $Mg_2SiO_4$ Film*

The amorphous $Mg_2SiO_4$ film was prepared on an aluminum substrate by pulsed laser ablation in the main chamber. During sample preparation, an $Mg_2SiO_4$ target was placed near the aluminum substrate at an approximately 40 mm distance and a 45° angle. The rotating target was irradiated with a pulsed laser beam (532 nm wavelength, 8 ns pulse width, 10 Hz repetition rate, and 20 mJ pulse energy) focused with an $f$ = 300 mm lens. Typically, laser ablation was performed for 12,000 pulses, resulting in a few tens of nanometers thick film on the aluminum substrate.

To characterize the samples by transmission electron microscopy (TEM), a thin film of $Mg_2SiO_4$ was prepared on a thin carbon film with a thickness of less than 6 nm on a copper mesh grid for TEM observation. A TEM apparatus (JEOL, JEM-2100F) was used with a field emission gun at an acceleration voltage of 200 kV, which was equipped with energy dispersive X-ray spectroscopy (EDS) for chemical composition analysis (JEOL, JED2300T). The TEM analysis results are summarized in Figure A1. The bright-field image (Figure A1a) shows that the sample was composed of ~99.7 % homogeneous regions and ~0.3 % high contrast particles by area ratio. The Mg/Si ratio of the homogeneous regions was analyzed by EDS and found to be 1.86. We obtained the electron diffraction patterns of a homogeneous region using a thicker sample fabricated by the same method (Figure A1b). The electron diffraction pattern shows a halo pattern that represents an amorphous structure. The intensity line profile of the electron diffraction pattern (Figure A1b, inset) shows a hump, which probably originates from a peak (2.88 Å) due to amorphous $Mg_2SiO_4$ (Igami et al. 2020).

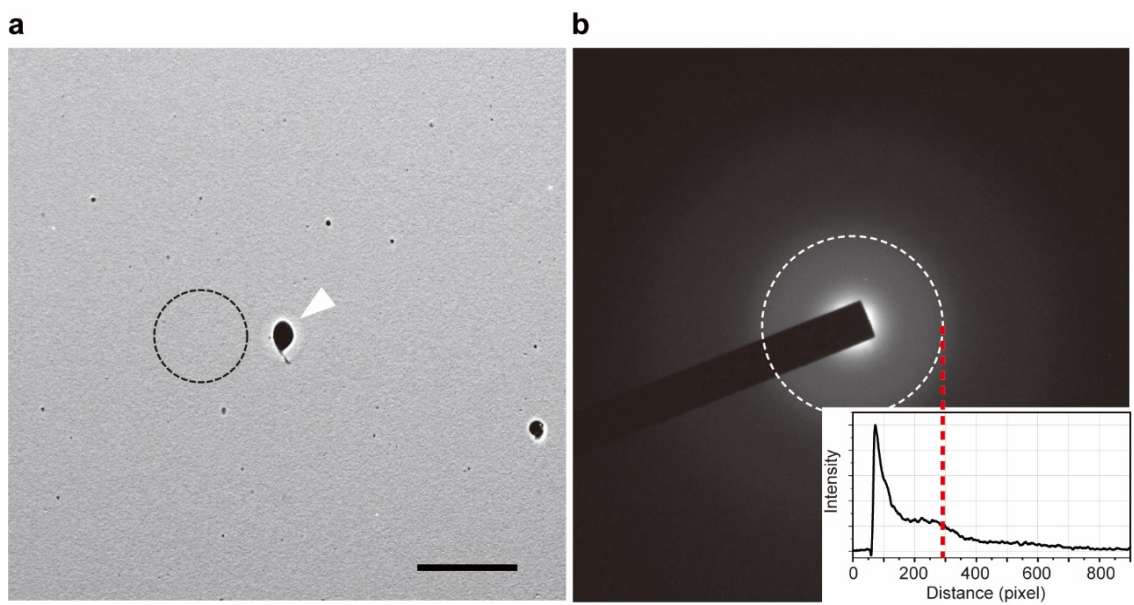

Figure A1. TEM analysis of amorphous $Mg_2SiO_4$ film. a, Bright-field TEM image of the amorphous $Mg_2SiO_4$ film. The region enclosed by the black broken line and the particle indicated by white arrowhead were analyzed by EDS, and the Mg/Si ratios are 1.86 and 1.74, respectively. The scale bar is 2 μm. b, Electron diffraction pattern of a homogeneous region in the amorphous $Mg_2SiO_4$ film. The white broken line corresponds to a d-spacing of 2.88 Å. The inset shows the normalized intensity line profile between the center of the diffraction pattern and the edge in the horizontal direction. The horizontal axis of the inset corresponds to the scale of the diffraction pattern along the horizontal direction.

The surface adsorption site density of the amorphous $Mg_2SiO_4$ film was estimated to be ~$10^{16}$ sites $cm^{-2}$ according to the method described elsewhere (Hama et al. 2012; Tsuge et al. 2019), in which the known site density of an Al surface ($1.2 \times 10^{15}$ sites $cm^{-2}$, Wyckoff 1931) was used as a reference. This value is similar to that of the porous ASW surface, indicating the roughness of the film.

*A.2. The effect of NSC during TPD process*

Because in the TPD-REMPI measurements, we monitor the number density of $o$-$H_2$ upon sublimation, i.e., at $t = t_{w.t.} + t_{TPD}$ defined in Figure 1, the measured number density of $o$-$H_2$ is different from that at $t = t_{w.t.}$. From Equation (1), the number density of $o$-$H_2$ after a certain waiting time ($t = t_{w.t.}$) is derived as $[o\text{-}H_2]_{t=t_{w.t.}} = [o\text{-}H_2]_0 e^{-k_{OP} t_{w.t.}}$. During a TPD run, ortho-to-para conversion is accelerated. When we ignore the temperature (time) dependence of the conversion rate during TPD, the time evolution of $[o\text{-}H_2]$ is described as

$$[o\text{-}H_2]_{t=t_{w.t.}+t_{TPD}} = [o\text{-}H_2]_0 \times e^{-k_{OP} t_{w.t.} - k' t_{TPD}}, \quad (A1)$$

where $k'$ is the ortho-to-para conversion rate constant during TPD. By taking the logarithm of Equation (A1), we obtain

$$\ln\bigl([o\text{-}H_2]_{t=t_{w.t.}+t_{TPD}}/[o\text{-}H_2]_0\bigr) = -k_{OP} t_{w.t.} - k' t_{TPD}, \quad (A2)$$

This relation suggests that a plot of $\ln\bigl([o\text{-}H_2]_{t=t_{w.t.}+t_{TPD}}/[o\text{-}H_2]_0\bigr)$ versus $t_{w.t.}$ gives a linear relation, and the rate constants $k_{OP}$ and $k'$ can be determined from the slope and intercept, respectively.

The experimental data obtained at 10 K (those shown in Figure 2b) are plotted in Figure A2. As expected, a plot of $\ln\bigl([o\text{-}H_2]_{t=t_{w.t.}+t_{TPD}}/[o\text{-}H_2]_0\bigr)$ versus $t_{w.t.}$ shows a linear relation, and from the slope and intercept, we obtain $1/k_{OP} = 1050 \pm 90$ s and $1/k' = 220 \pm 50$ s. The $1/k_{OP}$ value ($1050 \pm 90$ s) agrees well with the time constant ($980 \pm 90$ s) determined from a single exponential fitting to $[o\text{-}H_2]_{t=t_{w.t.}+t_{TPD}}$, indicating that the latter value can be regarded as the time constant for ortho-to-para conversion during the waiting time. In addition, the ortho-to-para conversion time constant during the TPD run, $1/k' = 220 \pm 50$ s, which is slightly smaller than or equivalent to the constant determined for 18 K ($260 \pm 70$ s), guarantees the validity of our analyses.

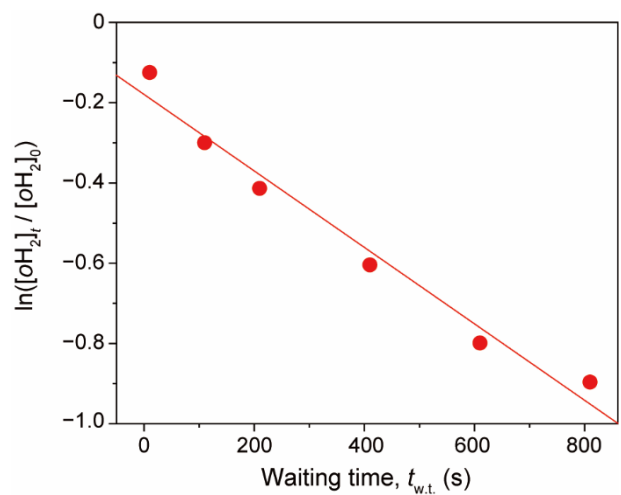

Figure A2. Additional analysis of the $J = 1$ TPD-REMPI signal decay at 10 K. The values $\ln\left([o\text{-}H_2]_{t=t_{w.t.}+t_{TPD}}/[o\text{-}H_2]_0\right)$ are plotted versus $t_{w.t.}$ according to Equation (A2). The linear regression analysis result is shown by the solid line.

*A.3. Para-to-ortho conversion of $D_2$ on a silicate surface*

Because $D_2$ is a Boson due to two deuterons with nuclear spin 1, the nuclear spin state of 2 and 0 (ortho) and 1 (para) are only coupled with the rotational states of even and odd quantum numbers, respectively. Thus, at low temperatures, *ortho*- and *para*-$D_2$ are in the $J = 0$ and 1 states, respectively.

The para-to-ortho conversion of $D_2$ molecules on a silicate surface was investigated at 18 K. The time variation of integrated TPD-REMPI signals are plotted in Figure A3. The sum of $J = 0$ and $J = 1$ signals was constant within experimental error, ensuring that desorption of $D_2$ by thermal and upon NSC was negligible. Fitting of the $J = 1$ signal by a single exponential function gave a decay time constant of $2{,}200 \pm 300$ s. The NSC of $D_2$ is much slower than $H_2$, similarly to those on the ASW surface as previously reported (Sugimoto & Fukutani 2011; Ueta et al. 2016).

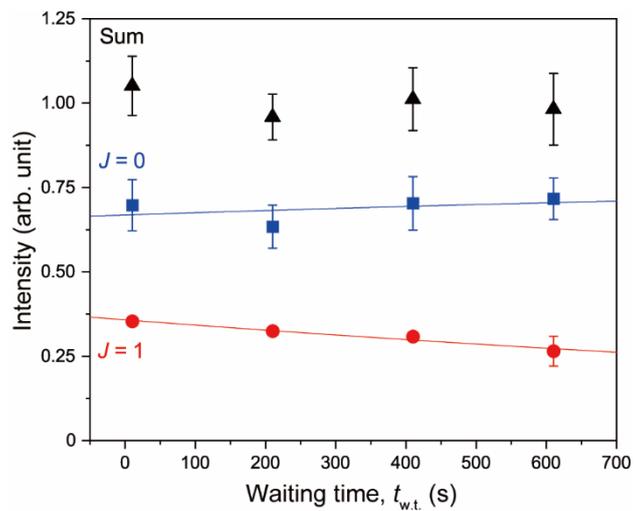

Figure A3. Time evolution of *ortho*-$D_2$ ($J = 0$) and *para*-$D_2$ ($J = 1$) populations at 18 K. The time evolution of TPD-REMPI signals for $J = 0$ (blue square), $J = 1$ (red circle), and their sum (black triangle). Intensities were obtained by integrating the TPD-REMPI signal. The error bars represent the statistical errors among 3 respective measurements. Solid curves are results of single exponential fitting.